\documentclass[prd,showpacs,preprintnumbers,twocolumn]{revtex4}
\usepackage{amsmath}
\usepackage{graphicx}
\usepackage{amsfonts}
\usepackage{amssymb}
\begin{document}

\title{Resonant interaction between gravitational waves, \\ electromagnetic waves and
plasma flows}
\author{Martin\ Servin and Gert\ Brodin\\Department of Physics, Ume\aa\ University,
S-901 87 Ume\aa, Sweden \\
Accepted for publication in Physical Review D\\
Scheduled issue: 15 August 2003.}
\begin{abstract}
In magnetized plasmas gravitational and electromagnetic waves may interact
coherently and exchange energy between themselves and with plasma flows. We
derive the wave interaction equations for these processes in the case of waves
propagating perpendicular or parallel to the plasma background
magnetic field. In the latter case, the electromagnetic waves are taken to be
circularly polarized waves of arbitrary amplitude. We allow for a background
drift flow of the plasma components which increases the number of possible
evolution scenarios. The interaction equations are solved analytically and the
characteristic time scales for conversion between gravitational and
electromagnetic waves are found. In particular, it is shown that in the
presence of a drift flow there are explosive instabilities resulting in the
generation of gravitational and electromagnetic waves. Conversely, we show
that energetic waves can interact to accelerate particles and thereby
\emph{produce} a drift flow. The relevance of these results for astrophysical
and cosmological plasmas is discussed.
\end{abstract}
\maketitle

\section{Introduction}

In empty and flat space-time linear gravitational wave ({\small GW})
perturbations and electromagnetic waves ({\small EMWs}) do not interact when
propagating in the same direction. If, however, there is a background
electromagnetic field or a background curvature present the two waves may
couple, interact and exchange energy \cite{Gertsenstein}-\cite{Grishchuk}.

Although {\small GW}s typically interact weakly with matter, it is of interest
to consider how {\small GW}-{\small EMW} interaction may be altered by the
presence of a magnetized plasma -- which is a common state of matter in
astrophysical and cosmological scenarios. The interaction may; lead to
production or modification of observable {\small EMW}s, as considered in the
Refs. \cite{Radiowave}-\cite{Phacc}, energize the plasma by exciting
Alfv\'{e}n and magnetosonic waves \cite{Alfvénwave}-\cite{Papadopulos} which
may be of importance in more complex processes such as supernovae explosions,
as discussed in Ref. \cite{SN}. Furthermore, it may even modify the expected
{\small GW} signals \cite{Kleidis} that are currently under attempted
detection and will, hopefully, provides us with a new window through which the
universe can be observed. Studying \emph{nonlinearly} interacting waves may
also reveal new types of instabilities that cannot be found using conventional
\emph{linear} stability theory, see e.g. Ref. \cite{Weiland}.

Wave interactions are most efficient if they are resonant (coherent), i.e. if
the frequencies satisfy certain matching conditions and the relative wave
phase remains unchanged for a long time. We will thus not consider incoherent
(non-resonant) wave interaction, in which case energy conversion takes place
on a much longer time scale. Resonant interaction in vacuum or in the presence
of a background electromagnetic field requires only that the frequencies
should coincide since both {\small GW}s and {\small EMW}s propagate along null
geodesics and therefore have identical phase velocity equal to $c$, the speed
of light. Considering {\small GW}-{\small EMW} interaction in a ``medium'' the
occurrence of such resonances is more rare because the phase velocity for
{\small GW} is (in the high frequency approximation and to a very high
accuracy) still equal to $c$, whereas for {\small EMW}s this occurs only for
particular wave frequencies (or in the limit of very high frequencies). For
this reason resonant {\small GW}-{\small EMW} interaction has been studied in
Ref. \cite{Alfvénwave} for the case of multiple {\small EMW}s, where two or
more {\small EMW}s -- that seperately do not propagate along null geodesics --
together produce perturbations that are resonant with a {\small GW}. See also
Ref. \cite{Anile}, where generation of {\small GW}s with the same mechanism
was considered but for interacting sound waves.

Ref. \cite{Greco} showed that resonant interaction between a {\small GW} and a
single low-frequency {\small EMW} (a ''magnetosonic wave'', in the terminology
of their formalism) can be realized for ``incompressible'' relativistic
magnetofluids, i.e. with sound velocity equal to $c$, but no calculations
regarding the coupling strength where made. In Ref. \cite{Papadopulos}\ the
excitation of magnetosonic waves by {\small GW}s was considered with
particular focus on almost coherent waves. In Ref. \cite{Macedo} kinetic
theory was used to derive, among other things, dispersion relations governing
the coupling between a {\small GW} and an {\small EMW} propagating
perpendicular to a background magnetic field. From the dispersion relations in
Ref. \cite{Macedo} it is clear that the waves are resonant if the wave
frequencies coincide with the electron plasma frequency (defined below) but
the strength of the interaction cannot easily be deduced from those results.

In this paper we study the interaction between {\small GW}s of small amplitude
and {\small EMW}s in a magnetized plasma modelled by multifluid equations. In
the case of perpendicular propagation with respect to the background magnetic
field, we take the {\small EMW} to be a \emph{high-frequency extraordinary
electromagnetic wave} (using standard terminology of plasma wave theory), the
only wave in this case for which we may have phase velocity equal to $c$
(without introducing a more complicated background state). The main focus of
our study is on the case of parallel propagation, where we make use of an
\emph{exact} (when neglecting gravitational effects) {\small EMW} solution
that can be reduced to e.g. high-frequency electromagnetic waves, whistler
waves, low-frequency Alfv\'{e}n waves or waves in electron-positron plasmas.
This solution also allows the presence of a relative drift flow of the fluids
constituting the plasma. The inclusion of a drift in the background state is
important for at least three reasons, when considering {\small GW}%
-{\small EMW} perturbations. Firstly, it increases the number of ways that
resonant interaction can occur, secondly, it alters the coupling strength
between the waves and, most importantly, it supplies the system with free
energy. This later fact means, as we will demonstrate, that the background
configuration may be unstable, leading to simultaneous generation of
{\small GW}s and {\small EMW}s.

The paper is organized as follows. In section II we give an overview of
certain principles of resonant wave interactions. The basic equations are
presented in section III, i.e. the Einstein-Maxwell system together with
multifluid equations. We demonstrate in section IV that, provided the
conditions for the high-frequency approximation are fulfilled, {\small GW}s
can be taken to be in the transverse traceless gauge also in the presence of matter and we
derive the corresponding evolution equations, describing how the {\small GW}s
are coupled to the matter perturbation. In section V we derive evolution
equations for the {\small EMW}s, including the effect of {\small GW}s on the
{\small EMW}s and expressions for the wave energy density. It is shown that
circularly polarized {\small EMW}s can have negative wave energy for certain
background parameter values and simultaneously be resonant with a {\small GW}.
Resonant interaction between the {\small GW}s and {\small EMW}s is studied in
section VI. Wave interaction equations are derived and their solutions are
examined, whereby the characteristic time scales for conversion between
gravitational and electromagnetic waves are found. For the case of negative
energy {\small EMW}s the coupling to {\small GW}s give rise to explosive
{\small EMW}-{\small GW} instabilities or, with different initial conditions,
acceleration of plasma flow. The results are summarized and further discussed
in section VII, including the relevance for astrophysical and cosmological plasmas.

\section{Overview of resonant wave interactions}

The natural wave modes of the gravitational field in the high-frequency
approximation \cite{Isacson} and of the electromagnetic field in magnetized
plasmas \cite{Plasmawaves} are well-known. Based on these solutions, we choose
a set of wave perturbations, $F_{n}$, that we represent with the following
complex notation: $F_{n}=f_{n}+c.c.$, where $f_{n}=\frak{f}_{n}e^{i\theta
_{(i)}}$, $\frak{f}_{n}$ is the complex amplitude, $\theta_{n}$ the (real)
wave phase and $c.c.$ stands for complex conjugation of the preceding term.
Letting $f_{1}$ denote the {\small GW} (e.g. some linear combination of
components of the perturbed metric tensor) and $f_{2}$, say, the electric
field of the {\small EMW}. From the governing equations, the
Einstein-Maxwell-plasma fluid system presented in section \ref{Basic eqs}, we
derive \emph{evolution equations} for these wave perturbations. As can be
found from section \ref{GW evol} and \ref{EM evol}, these equations take the
form
\begin{align}
\hat{D}_{1}f_{1} &  =S_{1}(f_{2},f_{1})+...\label{int_evol1}\\
\hat{D}_{2}f_{2} &  =S_{2}(f_{1},f_{2})+...\label{int_evol2}%
\end{align}
where $\hat{D}_{n}$ is the linear wave propagator and the interaction
\emph{source term} $S_{n}(f_{m},f_{n})$ is an algebraic expression
\cite{Algebraic} involving $f_{m}$ (and possibly also $f_{n}$) such that
$S_{n}\equiv\frak{S}_{n}e^{i\phi_{n}}$ is resonant with $f_{n}$, i.e. the
relative phase $\theta_{n}-\phi_{n}$ is constant. The dots in Eqs.
(\ref{int_evol1}) and (\ref{int_evol2}) indicates that there are in general
also terms that are non-resonant with $f_{n}$. Non-resonant terms are
discarded, as they only have an effect on a much larger time-scale. The source
term $S_{1}$ have the structure such that in the absence of {\small EMW}s,
$f_{2}=0$, $S_{1}$ vanishes and Eq. (\ref{int_evol1}) reduces to the free wave
equation for {\small GW}s, $[\partial_{t}^{2}-\partial_{z}^{2}]f_{1}=0$.
Conversely, if there are no {\small GW}s then $S_{2}$ vanishes and Eq.
(\ref{int_evol2}) reduces to the plasma {\small EMW} equation that corresponds
to the assumed perturbation.

Our immediate purpose is to determine the coupling strength and thereby the
characteristic time-scales for the wave interactions. We do this by studying
spatially uniform monochromatic waves propagating in a (locally) static and
uniform background (the relevance for less idealized situations will be
discussed in Section \ref{Summary disc}). The interaction will then result in
time dependent wave amplitudes, governed by the derived evolution equations.
Since the source terms are small -- either proportional to the gravitational
coupling constant or to the small {\small GW} amplitude -- we assume that
$|\omega_{n}\frak{f}_{n}|\gg|\partial_{t}\frak{f}_{n}|$. This means that, to
lowest order, the waves will propagate freely according to the dispersion
relation $D_{n}(-i\omega_{n},i\mathbf{k}_{n})=0$ and, to the next level of
accuracy, have a slowly evolving amplitude. Applying this, the differential
operators in Eqs. (\ref{int_evol1}) and (\ref{int_evol2}) can be approximated
by $\hat{D}\approx D(-i\omega,i\mathbf{k})+(\partial_{\omega}D)\tilde
{\partial}_{t}$, where the tilde notation on $\tilde{\partial}_{t}$ means that
the derivative acts only on the amplitude. Higher order derivatives are
neglected. The evolution equations (\ref{int_evol1}) and (\ref{int_evol2})
thus becomes
\begin{align}
\tilde{\partial}_{t}f_{1} &  =\left(  \partial_{\omega_{1}}D_{1}\right)
^{-1}S_{1}(f_{2},f_{1})\label{int_int1}\\
\tilde{\partial}_{t}f_{2} &  =\left(  \partial_{\omega_{2}}D_{2}\right)
^{-1}S_{2}(f_{1},f_{2})\label{int_int2}%
\end{align}
and we refer to these as the \emph{interaction equations}. Determining $S_{1}$
and $S_{2}$ and studying the solutions of these equations are the main
purposes of this paper.

\section{Basic equations\label{Basic eqs}}

We take the gravitational and electromagnetic field to be governed by the
Einstein field equations
\begin{equation}
G_{ab}=\kappa T_{ab}\label{EFE}%
\end{equation}
and the Maxwell equations
\begin{align}
\nabla_{a}F^{ab} &  =j^{b}\label{ME1}\\
\nabla_{a}F_{bc}+\nabla_{b}F_{ca}+\nabla_{c}F_{ab} &  =0\label{ME2}%
\end{align}
where $G_{ab}$ is the Einstein tensor, $\kappa=8\pi G$, $T_{ab}$ is the
energy-momentum tensor, $F_{ab}$ is the electromagnetic field tensor, $j^{a}$
is the total four-current density and $\nabla$ denotes covariant derivative.
We use units where the velocity of light in vacuum is $c=1$. The metric
signature is $(-+++)$ and for the indices we use $a,b,c,...=0,1,2,3$ and
$i,j,k,...=1,2,3$. The matter present in the interaction region is a
magnetized plasma for which we choose a multifluid description. This means
that we take the plasma to consist of a number of interpenetrating charged
fluids, one for each species of particles that constitutes the plasma. The
fluids interact only through the electromagnetic and gravitational field, i.e.
we neglect effects of particle collisions as can be done for most plasmas. The
appropriate fluid equations are then, for each fluid component $(i)$, the
equation of mass conservation
\begin{equation}
\nabla_{a}(m_{(i)}n_{(i)}u_{(i)}^{a})=0\label{cont}%
\end{equation}
and the momentum equation
\begin{equation}
m_{(i)}n_{(i)}u_{(i)}^{a}\nabla_{a}u_{(i)}^{b}=-\nabla^{b}p_{(i)}%
+q_{(i)}n_{(i)}u_{(i)}^{a}F_{a}^{b}\label{momentum}%
\end{equation}
where $n_{(i)}$ is the proper particle number density, $u_{(i)}^{a}$ is the
fluid four-velocity, $m_{(i)}$ is the particle mass, $p_{(i)}$ is the pressure
and $q_{(i)}$ is the particle electric charge. For closure this should be
supplemented by some equation of state which we do not specify here. The total
current density is then given by $j^{a}=\sum_{(i)}q_{(i)}n_{(i)}u_{(i)}^{a}$
and the energy-momentum tensor by $T_{ab}=T_{ab}^{(fl)}+T_{ab}^{(em)}$, where
the fluid contribution is
\begin{equation}
T_{ab}^{(fl)}=\sum_{(i)}(m_{(i)}n_{(i)}+p_{(i)})u_{(i)a}u_{(i)b}+p_{(i)}%
g_{ab}\label{Tfl}%
\end{equation}
and the electromagnetic field contribution is
\begin{equation}
T_{ab}^{(em)}=F_{a}^{\;c}F_{bc}-{\tfrac{1}{4}}g_{ab}F^{cd}F_{cd}\label{Tem}%
\end{equation}
From now on we will omit the species index, $(i)$, unless there can be any confusion.

It is practical to rewrite the Maxwell and fluid equations in a more
convenient form. In particular we intend to introduce an orthonormal frame so
that the interpretation of the quantities becomes more direct. First, we note
that by introducing an observer four velocity, $V^{a}$, the electromagnetic
field can be decomposed, relative to $V^{a}$, into an electric part, $E^{a}$,
and a magnetic part, $B^{a}$, according to $F^{ab}=V^{a}E^{b}-V^{b}%
E^{a}+\epsilon^{abc}B_{c}$, where $E_{a}=F_{ab}V^{b}$, $B_{a}=\frac{1}%
{2}\epsilon^{abc}F_{bc}$, $\epsilon_{abc}=V^{d}\epsilon_{abcd}$ and
$\epsilon_{abcd}$ is the 4 dimensional volume element with $\epsilon
_{0123}=\sqrt{|\det g|}$. If one chooses an orthonormal frame (in which
$V^{a}=(1,\mathbf{0})$) with contravariant basis $\{\mathbf{e}_{a}\}$, the
Maxwell and fluid equations can be formulated as \cite{Radiowave},
\cite{source terms}%
\begin{align}
\mathbf{\nabla\cdot E} &  =\rho+\rho_{E}\label{plasma1}\\
\mathbf{\nabla\cdot B} &  =\rho_{B}\label{plasma2}\\
\mathbf{e}_{0}\mathbf{E}-\mathbf{\nabla\times B} &  =-\mathbf{j}%
-\mathbf{j}_{E}\label{plasma3}\\
\mathbf{e}_{0}\mathbf{B}+\mathbf{\nabla\times E} &  =-\mathbf{j}%
_{B}\label{plasma4}\\
\mathbf{e}_{0}(\gamma n)+\mathbf{\nabla}\cdot(\gamma n\mathbf{v}) &  =\Delta
n\label{plasma5}\\
mn(\mathbf{e}_{0}+\mathbf{v}\cdot\mathbf{\nabla)\gamma v} &  =-\gamma
^{-1}\mathbf{\nabla}p\nonumber\\
&  +qn(\mathbf{E}+\mathbf{v}\times\mathbf{B})+mn\mathbf{g}\label{plasma6}%
\end{align}
where we have introduced an Euclidean three-vector notation $\mathbf{E}%
=(E^{1},E^{2},E^{3})$ etc., $\mathbf{\nabla}=(\mathbf{e}_{1},\mathbf{e}%
_{2},\mathbf{e}_{3})$, $\rho=j^{0}$ and coordinate velocity, $v^{a}$, so that
$u^{a}=\gamma v^{a}$ with $\gamma=(1-v_{i}v^{i})^{-\frac{1}{2}}$. The effect
of the gravitational field is included in the form of \emph{effective}
charges, currents, gravitational forces etc. that were introduced above, and
they are given by \cite{source terms}%
\begin{align*}
\rho_{E} &  \equiv-\Gamma_{\,ji}^{i}E^{j}-\epsilon^{ijk}\Gamma_{\,ij}^{0}%
B_{k}\\
\rho_{B} &  \equiv-\Gamma_{\,ji}^{i}B^{j}+\epsilon^{ijk}\Gamma_{\,ij}^{0}%
E_{k}\\
\mathbf{j}_{E} &  \equiv\left[  -\left(  \Gamma_{\,0j}^{i}-\Gamma_{\,j0}%
^{i}\right)  E^{j}+\Gamma_{\,0j}^{j}E^{i}\right.  \\
&  \left.  -\epsilon^{ijk}\left(  \Gamma_{\,j0}^{0}B_{k}+\Gamma_{\,jk}%
^{m}B_{m}\right)  \right]  \mathbf{e}_{i}\\
\mathbf{j}_{B} &  \equiv\left[  -\left(  \Gamma_{\,0j}^{i}-\Gamma_{\,j0}%
^{i}\right)  B^{j}+\Gamma_{\,0j}^{j}B^{i}\right.  \\
&  \left.  +\epsilon^{ijk}\left(  \Gamma_{\,j0}^{0}E_{k}+\Gamma_{\,jk}%
^{m}E_{m}\right)  \right]  \mathbf{e}_{i}\\
\Delta n &  \equiv-\gamma n\left(  \Gamma_{\,0i}^{i}+\Gamma_{\,00}^{i}%
v_{i}+\Gamma_{\,ji}^{i}v^{j}\right)  \\
\mathbf{g} &  \equiv-\gamma\left[  \Gamma_{\,00}^{i}+\left(  \Gamma_{\,0j}%
^{i}+\Gamma_{\,j0}^{i}\right)  v^{j}+\Gamma_{\,jk}^{i}v^{j}v^{k}\right] \mathbf{e}_{i}
\end{align*}
where $\Gamma_{\,bc}^{a}$ are the Ricci rotation coefficients.

\section{Gravitational wave evolution}

\label{GW evol}In this section we show that, in the high-frequency
approximation \cite{Isacson} for {\small GW}s propagating in matter, the
{\small GW}s can be taken to be in the transverse trace-less (TT) gauge and we
give the corresponding evolution equation. In flat space-time linearized
gravitational waves can always be taken to be in the TT gauge. In vacuum
space-time with a background curvature this is in general only possible in the
high frequency limit, where the ratio of the {\small GW} wavelength and the
characteristic background length scale (radius of curvature) tends to zero.
The advantages of having {\small GW}s in the TT-gauge is that the evolution
equation becomes more simple, the polarization state is more clear and, most
important for us here, it reduces the amount of algebra when calculating the
effect on a plasma, e.g. when calculating the gravitational source terms in
the {\small EMW} evolution equation. However, the (perturbed) energy-momentum
tensor should have the same properties as the perturbed Einstein tensor and
this will in general not be the case if one just assumes the TT-gauge.
Nevertheless, we demonstrate that the TT gauge can also be applied in the
presence of matter provided the conditions for the high frequency
approximation is fulfilled. This result seems to have been overlooked in the litterature.

Let the background gravitational field and the unperturbed plasma fulfill the
Einstein field equations, $G_{ab}^{(0)}=\kappa T_{ab}^{(0)}$, and introduce a
perturbation so that $G_{ab}=G_{ab}^{(0)}+\delta G_{ab}$ and $T_{ab}%
=T_{ab}^{(0)}+\delta T_{ab}$. We assume the high frequency approximation so
that the background space-time can be taken to be Minkowski space-time
\cite{Isacson} and put $g_{ab}=\eta_{ab}+h_{ab}$, where $\eta_{ab}$ is the
Minkowski metric and $h_{ab}$ the small metric perturbation that should obey
\begin{equation}
\delta G_{ab}=\kappa\delta T_{ab}\label{pertEFE}%
\end{equation}
We limit ourself to linear gravitational perturbations. The perturbed Einstein
tensor, linearized in $h_{ab}$, is given by
\begin{equation}
\delta G_{ab}[\bar{h}]=-\tfrac{1}{2}\partial^{c}\partial_{c}\bar{h}%
_{ab}+\partial^{c}\partial_{(b}\bar{h}_{a)c}-\tfrac{1}{2}\eta_{ab}\partial
^{c}\partial^{d}\bar{h}_{cd}\label{dG}%
\end{equation}
$\bar{h}_{ab}\equiv h_{ab}-\frac{1}{2}\eta_{ab}h$ and the brackets $_{(}$
$_{)}$ stands for symmetrization with respect to the enclosed indices. In
vacuum, the next step would be to apply the Lorentz gauge condition,
$\partial^{b}\bar{h}_{ab}=0$. Instead of this, we split $\bar{h}_{ab}$
according to
\begin{equation}
\bar{h}_{ab}=\bar{h}_{ab}^{L}+\bar{f}_{ab}\label{metric split}%
\end{equation}
where $\bar{h}_{ab}^{L}$ is the (maximal) part of $\bar{h}_{ab}$ that fulfills
the Lorentz gauge condition and the other part, $\bar{f}_{ab}$, containing
terms that \emph{cannot} be fitted into $\bar{h}_{ab}^{L}$ (both terms should
be symmetric). With this splitting Eq. (\ref{pertEFE}) reads
\begin{equation}
-\tfrac{1}{2}\partial^{c}\partial_{c}\bar{h}_{ab}^{L}+\delta G_{ab}[\bar
{f}]=\kappa\delta T_{ab}\label{splitEFE}%
\end{equation}
Next we assume that the metric perturbation (and the perturbed energy-momentum
tensor) is of the form $h_{ab}=\frak{h}_{ab}(x^{d})e^{ik_{c}x^{c}}+c.c.$ with
the dispersion relation $k_{a}k^{a}=0$. The dependency of $\frak{h}_{ab}$ on
$x^{d}$ is assumed weak, i.e. $|\partial_{c}\frak{h}_{ab}|\ll|k_{c}%
\frak{h}_{ab}|$. Computing $\delta G_{ab}[\bar{f}]$ gives to lowest order
\begin{equation}
\delta G_{ab}[\bar{f}]=-k^{c}k_{(b}\bar{f}_{a)c}+\tfrac{1}{2}\eta_{ab}%
k^{c}k^{d}\bar{f}_{cd}\label{dGf}%
\end{equation}
where terms involving slow derivatives (derivatives on the weakly varying
amplitude) have been neglected. From this expression it follows, for
perturbations with four wave vector $k^{a}=(\omega,0,0,\omega)$, that $\delta
G_{12}[\bar{f}]=\delta G_{21}[\bar{f}]=0$ and $\delta G_{11}[\bar{f}]=\delta
G_{22}[\bar{f}]$. Furthermore, nonzero components of $\bar{f}_{ab}$ should be
of order $\bar{f}\sim\kappa\delta T/\omega^{2}$ (which must be much smaller
than unity in order not to invalidate the linearization in $h_{ab}$) for the
field equations (\ref{splitEFE}) to be fulfilled.

We now turn to the terms in $\bar{h}_{ab}^{L}$. A gauge transformation
$x^{a^{\prime}}=x^{a}+\xi^{a}$ with $\xi^{a}\ll1$ and $\xi^{a}=-i\tilde{\xi
}^{a}(x^{c})e^{ik_{b}x^{b}}+c.c.$ alters the metric perturbation by
\begin{equation}
\bar{h}^{a^{\prime}b^{\prime}}=\bar{h}^{ab}-\tilde{\xi}^{a}k^{b}-\tilde{\xi
}^{b}k^{a}+\eta^{ab}\tilde{\xi}^{c}k_{c} \label{hprim}%
\end{equation}
where it is understood that the amplitude, $\tilde{\xi}^{a}(x^{c})$, is weakly
dependent on $x^{c}$ in the same way as $\bar{h}^{ab}$. We can, however,
neglect slow derivatives on $\xi^{a}$ as such terms only produce small
corrections to the $\bar{f}_{ab}$ part or, finally in the wave evolution
equations, terms of second order slow derivatives. Consequently, $\xi^{a}$
can, as in the vacuum case, be chosen as linear combinations of $\bar{h}%
_{ab}^{L}$ such that
\begin{equation}
\bar{h}^{a^{\prime}b^{\prime}}=\bar{h}_{TT}^{ab}+\bar{f}^{ab} \label{hprimTT}%
\end{equation}
where $\bar{h}_{TT}^{ab}$ is ``in the transverse traceless gauge'', i.e.
$\bar{h}_{TT}^{11}=-\bar{h}_{TT}^{22}\equiv h_{A}$ and $\bar{h}_{TT}^{12}%
=\bar{h}_{TT}^{21}\equiv h_{B}$. The evolution equations for $h_{A}$ and
$h_{B}$ follows from Eq. (\ref{splitEFE}) by subtracting the $_{11}$ and the
$_{22}$ components and by -- for aestethical reasons only -- adding the
$_{12}$ and $_{21}$ components. The result is
\begin{align}
\partial^{c}\partial_{c}h_{A}  &  =-\kappa\left(  \delta T_{11}-\delta
T_{22}\right) \label{hA}\\
\partial^{c}\partial_{c}h_{B}  &  =-\kappa\left(  \delta T_{12}+\delta
T_{21}\right)  \label{hB}%
\end{align}
which depends critically on the fact that $\delta G_{12}[\bar{f}]=\delta
G_{21}[\bar{f}]=0$ and $\delta G_{11}[\bar{f}]=\delta G_{22}[\bar{f}]$.

To summarize, for the part of the metric perturbation that represents
{\small GW}s propagating in the $x^{3}$-direction, $\bar{h}_{TT}^{ab}$, we
have derived evolution equations that describes how the waves are coupled to
the perturbed energy-momentum tensor. Clearly, nonvanishing $\delta
T_{11}-\delta T_{22}$ and $\delta T_{12}$ can potentially act to drive or damp
{\small GW}s. The remaining part of the perturbed metric, given by $\bar
{f}_{ab}$, should be included for the sake of consistency. The perturbed
energy-momentum that is not accounted for in Eqs. (\ref{hA}) and (\ref{hB})
produces some gravitational response, which is given by $\delta G_{ab}[\bar
{f}]$ and found from the remaining components of Eqs. (\ref{splitEFE}) (due to
Eqs. (\ref{splitEFE}) and (\ref{dGf}) these are purely algebraic equations).
As $\bar{f}_{ab}$ represents the self-gravitation of the energy-momentum
perturbation it is in general important for the evolution of the matter and
electromagnetic field, but since $\bar{f}\sim\kappa\delta T/\omega^{2}$ it is
negligible when we consider waves in the high-frequency limit. We conclude
that in the high-frequency limit and for slowly evolving wave amplitudes the
{\small GW}s can be taken to be in the TT-gauge and the metric perturbation
$h_{ab}=h_{ab}^{TT}$ evolves according to Eqs. (\ref{hA}) and (\ref{hB}). In
this limit the energy density carried by {\small GW}s follows directly from
the Landau-Lifshitz pseudo energy-momentum tensor \cite{Landau}
\begin{equation}
\mathcal{E}_{{\small GW}}={\tfrac{1}{2\kappa}}\left[  (\partial_{t}h_{A}%
)^{2}+(\partial_{t}h_{B})^{2}\right]  \label{EGW}%
\end{equation}

When studying the effect of {\small GW}s on a magnetized plasma it is
practical to use tetrad formalism and introduce an orthonormal frame. A
contravariant basis that corresponds to {\small GW}s in the TT gauge with
propagation in the $x^{3}\equiv z$ direction is given by
\begin{align*}
\mathbf{e}_{0}  &  =\partial_{t}\;,\;\;\mathbf{e}_{1}=(1-{\tfrac{1}{2}}%
h_{A})\partial_{x}-{\tfrac{1}{2}}h_{B}\partial_{y}\;,\\
\mathbf{e}_{2}  &  =(1+{\tfrac{1}{2}}h_{A})\partial_{y}-{\tfrac{1}{2}}%
h_{B}\partial_{x}\;,\;\;\mathbf{e}_{3}=\partial_{z}%
\end{align*}
and the gravitationally induced terms in Eqs. (\ref{plasma1})-(\ref{plasma6})
are (displaying the nonzero terms only)
\begin{align}
\mathbf{g}  &  =-{\tfrac{1}{2}}\gamma(1-v_{z})\left[  v_{x}\partial_{t}%
h_{A}+v_{y}\partial_{t}h_{B})\right]  \mathbf{e}_{1}\nonumber\\
&  -{\tfrac{1}{2}}\gamma(1-v_{z})\left[  v_{x}\partial_{t}h_{B}-v_{y}%
\partial_{t}h_{A}\right]  \mathbf{e}_{2}\nonumber\\
&  -{\tfrac{1}{2}}\gamma\left[  (v_{x}^{2}-v_{y}^{2})\partial_{t}h_{A}%
+2v_{x}v_{y}\partial_{t}h_{B}\right]  \mathbf{e}_{3}\label{g}\\
\mathbf{j}_{E}  &  =-{\tfrac{1}{2}}\left[  (E_{x}-B_{y})\partial_{t}%
h_{A}+(E_{y}+B_{x})\partial_{t}h_{B}\right]  \mathbf{e}_{1}\nonumber\\
&  -{\tfrac{1}{2}}\left[  -(E_{y}+B_{x})\partial_{t}h_{A}+(E_{x}%
-B_{y})\partial_{t}h_{B}\right]  \mathbf{e}_{2}\label{jE}\\
\mathbf{j}_{B}  &  =-{\tfrac{1}{2}}\left[  (E_{y}+B_{x})\partial_{t}%
h_{A}-(E_{x}-B_{y})\partial_{t}h_{B}\right]  \mathbf{e}_{1}\nonumber\\
&  -{\tfrac{1}{2}}\left[  (E_{x}-B_{y})\partial_{t}h_{A}+(E_{y}+B_{x}%
)\partial_{t}h_{B}\right]  \mathbf{e}_{2} \label{jB}%
\end{align}
where we have used $\partial_{z}\approx-\partial_{t}$.

\section{Electromagnetic wave evolution\label{EM evol}}

In this section we present the evolution equations for the {\small EMW}s that
we are considering. Since the {\small GW} four-wave vector is assumed to
fulfill the dispersion relation $k_{a}k^{a}=0$, they have phase velocity equal
to unity.\ The {\small EMW}s that can interact resonantly with the
{\small GW}s should then also have phase velocity equal to unity. For
simplicity, we restrict our study to the case of perpendicular and parallel
propagation to the background magnetic field, respectively.

The plasma is assumed neutral and uniform, with a constant background magnetic
field, $B_{(0)}$. The plasma frequency is denoted by $\omega_{p(i)}%
\equiv(n_{(i)}q_{(i)}^{2}/m_{(i)})^{1/2}$ and the cyclotron frequency by
$\omega_{c(i)}\equiv q_{(i)}B_{(0)}/\gamma_{(i)}m_{(i)}$, where $n_{(i)}$ is
the unperturbed proper particle number density. We also allow the presence of
a background drift flow in the direction of the magnetic field. The inclusion
of the drift flow provides the system with free energy that may be transferred
to the {\small GW}s and {\small EMW}s during the interaction.

In magnetized plasmas there are many electromagnetic wave modes and their
properties are well known for flat space-time and for linear (small amplitude)
waves (see e.g. \cite{Plasmawaves}). In order to obtain the effect of
{\small GW}s on the {\small EMW}s we derive their evolution equations from the
basic equations (\ref{plasma1})-(\ref{plasma6}), including the effect of the
gravitational field. For later reference, we note that the total energy
density of an {\small EMW} is given by%
\begin{equation}
\mathcal{E}_{EM}\textsl{=}\mathbf{E}^{\ast}\cdot\left(  \omega^{-1}%
\partial_{\omega}\left[  \omega^{2}\epsilon\right]  \right)  \cdot
\mathbf{E}\label{EEM}%
\end{equation}
where $\mathbf{E}\propto e^{i(kz-\omega)t}$ is the electric field of the wave,
$\omega$ is the frequency and $\epsilon$ is the effective dielectric tensor
defined such that $\mathbf{D}=\epsilon\mathbf{\cdot E}$ gives the electric
displacement field. For the cold (i.e. zero pressure) linearized plasma fluid
equations $\mathbf{\epsilon}$ is given by\cite{Chen}
\begin{equation}
\mathbf{\epsilon=}\left(
\begin{array}
[c]{ccc}%
S & -iD & 0\\
iD & S & 0\\
0 & 0 & P
\end{array}
\right)  \label{dielectric}%
\end{equation}
where there is a background magnetic field and drift flow $v_{z}$ in the
$z$-direction, $S\equiv\frac{1}{2}\left(  R+L\right)  $, $D\equiv\frac{1}%
{2}\left(  R-L\right)  $ and
\begin{align*}
P &  \equiv1-\sum_{(i)}\frac{\omega_{p(i)}^{2}}{\left(  \omega-kv_{z(i)}%
\right)  ^{2}}\\
R &  \equiv1-\sum_{(i)}\frac{\omega_{p(i)}^{2}\left(  \omega-kv_{z(i)}\right)
}{\omega^{2}(\omega-kv_{z(i)}+\omega_{c(i)})}\\
L &  \equiv1-\sum_{(i)}\frac{\omega_{p(i)}^{2}\left(  \omega-kv_{z(i)}\right)
}{\omega^{2}(\omega-kv_{z(i)}-\omega_{c(i)})}%
\end{align*}
Although this result concerns linear waves in a cold plasma it also applies
for the large amplitude waves in a moderately warm plasma that we will
consider below. This is because they are circularly polarized and purely
transverse, making them effectively linear \cite{efflin} and with no density
(nor pressure) perturbations.

\subsection{Perpendicular propagation}

In the case of perpendicular propagation the plasma equations are difficult to
treat without resorting to perturbation theory. We restrict ourself in this
case to linear perturbations and to this level of approximation the wave
perturbations do not produce any effective force, $\mathbf{g}$, that is
\emph{not} rapidly oscillating. This implies that the waves cannot exchange
energy with the drift flow in an efficient way. Consequently, we discard the
drift flow entirely and only considers one particular wave mode in an
electron-ion plasma, namely the extraordinary electromagnetic wave -- using
standard plasma physics terminology. This is the only wave that can be
resonant with a {\small GW} in the simplest plasma model, the \emph{cold
plasma fluid model}.

We take the background magnetic field to be $\mathbf{B}_{(0)}=B_{(0)}%
\mathbf{e}_{1}$. The extraordinary {\small EMW} is a high-frequency wave,
$\omega\gg\omega_{pi}$ (such that only the electron fluid is perturbed), and
has, when propagating in the $z$-direction, the following non-zero linear
perturbations: $\delta n$, $\mathbf{v}=v_{y}\mathbf{e}_{2}+v_{z}\mathbf{e}%
_{3}$, $\mathbf{E}=E_{y}\mathbf{e}_{2}+E_{z}\mathbf{e}_{3}$ and $\delta
\mathbf{B}=\delta B_{x}\mathbf{e}_{1}$, where the density and velocity
perturbation referres to the electron fluid only. For a cold plasma, the Eqs.
(\ref{plasma1})-(\ref{plasma6}) reduces to the following equations, linearized
in the small perturbations
\begin{align}
\partial_{z}E_{z}  &  =q\delta n\label{Xeq1}\\
\partial_{t}E_{y}-\partial_{z}\delta B_{x}  &  =-qnv_{y}-\left.  j_{E}\right.
_{y}\\
\partial_{t}E_{z}  &  =-qnv_{z}\\
\partial_{t}\delta B_{x}-\partial_{z}E_{y}  &  =-\left.  j_{B}\right.  _{x}\\
\partial_{t}v_{y}  &  =\frac{q}{m}(E_{y}+v_{z}B_{(0)})\\
\partial_{t}v_{z}  &  =\frac{q}{m}(E_{z}-v_{y}B_{(0)})\\
\partial_{t}\delta n+n\partial_{z}v_{z}  &  =0 \label{Xeq7}%
\end{align}
which can be combined into the following evolution equation for the wave
magnetic field (after one time integration)
\begin{equation}
\hat{D}_{X}\delta B_{x}=S_{X} \label{Xwaveq}%
\end{equation}
where the wave propagator, $\hat{D}_{X}$, and the source term, $S_{X}$, are%
\[
\hat{D}_{X}\equiv\partial_{t}^{4}+(\omega_{p}^{2}+\omega_{h}^{2})\partial
_{t}^{2}-\partial_{t}^{2}\partial_{z}^{2}-\omega_{h}^{2}\partial_{z}%
^{2}+\omega_{p}^{4}%
\]%
\begin{align*}
S_{X}  &  \equiv{\tfrac{1}{2}}B_{(0)}\left[  \partial_{t}^{4}+(\omega_{p}%
^{2}+\omega_{h}^{2})\partial_{t}^{2}+\omega_{p}^{4}\right. \\
&  \left.  +\partial_{z}\partial_{t}(\partial_{t}^{2}+\omega_{p}^{2}%
+\omega_{c}^{2})\right]  h_{A}%
\end{align*}
and we have made use of Eqs. (\ref{g})-(\ref{jB}), linearized in the wave
perturbations and introduced the hybrid frequency $\omega_{h}^{2}\equiv
\omega_{p}^{2}+\omega_{c}^{2}$. In the absence of {\small GW}s Eq.
(\ref{Xwaveq}) reduces to $\hat{D}_{X}\delta B_{x}=0$ and, in particular, for
wave perturbations $\delta B_{x}\propto e^{i(kz-\omega t)}$ it produces the
dispersion relation for the extraordinary wave
\begin{equation}
D_{X}\equiv\omega^{4}-(\omega_{h}^{2}+\omega_{p}^{2}+k^{2})\omega^{2}%
+\omega_{h}^{2}k^{2}+\omega_{p}^{4}=0 \label{DX}%
\end{equation}
It should be pointed out that the extraordinary mode, with the given
polarization, only couples to $h_{A}$ and not to $h_{B}$. Computing the
perturbed energy-momentum tensor gives, linear in the perturbations,
\begin{align}
\delta T_{11}  &  =-\delta T_{22}=-\delta B_{x}B_{(0)}\label{dT11_B}\\
\delta T_{12}  &  =\delta T_{21}=0 \label{dT12_B}%
\end{align}
The dispersion relation (\ref{DX}) implies that the condition for having phase
velocity $\omega/k=1$ is $\omega=\omega_{p}$. We also note that from Eq.
(\ref{EEM}) and Eqs. (\ref{Xeq1})-(\ref{Xeq7}) it follows, after some algebra,
that the energy density of a free extraordinary wave with $\omega=\omega_{p}$
is given by%
\begin{equation}
\mathcal{E}_{X}=2\omega_{c}^{-2}\omega_{h}^{2}\left|  \delta B_{x}\right|
^{2} \label{EX}%
\end{equation}

\subsection{Parallel propagation}

If one neglects general relativistic effects, there exist exact solutions of
the multifluid equations that represents {\small EMW}s propagating parallel to
a background magnetic field (see e.g. \cite{Stenflo}). These solutions
describe circularly polarized waves of arbitrary amplitude and arbitrary
frequency and can thus, taking the appropriate limits, be reduced to
high-frequency electromagnetic waves, whistler waves, low-frequency Alfv\'{e}n
waves or waves in electron-positron plasmas. The solution can also be extended
to include a background drift flow \cite{Flow}. We derive the evolution
equation for this wave for a two-component plasma, including the effects of
{\small GW}s. The indices $e$ and $i$ will now stand for negatively charged
particles (e.g. electrons) and positively charged particles (e.g. positive
ions), respectively, but we make no assumptions for the mass ratio
$m_{e}/m_{i}$. The calculations are therfore equally valid for electron-ion
type of plasmas as well as for electron-positron type of plasmas.

We take the background magnetic field to be $\mathbf{B}_{(0)}=B_{(0)}%
\mathbf{e}_{3}$ and suppose there is a velocity drift, $v_{z}$, in the
$z$-direction. The above mentioned circularly polarized {\small EMW},
propagating in the $z$-direction, has the following perturbations (of
arbitrary amplitude) $\mathbf{v}=v_{x}\mathbf{e}_{1}+v_{y}\mathbf{e}_{2}$,
$\mathbf{E}=E_{x}\mathbf{e}_{1}+E_{y}\mathbf{e}_{2}$ and $\delta
\mathbf{B}=\delta B_{x}\mathbf{e}_{1}+\delta B_{y}\mathbf{e}_{2}$, where
$\mathbf{v}$ and $\mathbf{\delta B}$ are parallel and all perturbations are
functions of $z$ and $t$ alone. Note that circular polarization implies that
$\gamma$ depends only on the wave \emph{amplitude} and on $v_{z}$, not on the
rapidly varying phase. In the following we treat $\gamma$ as being constant
and $v_{z}$ as a constant uniform background flow. Small (linear) deviations
in $\gamma$ and $v_{z}$ can still be described (see Section \ref{Res wave int}
and \ref{Summary disc}). For convenience we introduce the variables $E_{\pm
}=E_{x}\pm iE_{y}$, and similarly for all other vector variables. In these
variables, suitable for circularly polarized waves (the plus/minus variables
corresponds to the amplitudes of the right/left hand polarization), those
equations of Eqs. (\ref{plasma1})-(\ref{plasma6}) that governs the given
perturbations can be rewritten as%

\begin{align}
(\partial_{t}+v_{z}\partial_{z}\mathbf{)}\gamma v_{\pm}  &  =\frac{q}%
{m}(E_{\pm}\mp iB_{0}v_{\pm}\pm iv_{z}B_{\pm})+g_{\pm}\label{Dvz}\\
\partial_{t}E_{\pm}  &  =\pm i\partial_{z}B_{\pm}-\Sigma qn\gamma v_{\pm
}-j_{E\pm}\label{DEpm}\\
\partial_{t}B_{\pm}  &  =\mp i\partial_{z}E_{\pm}-j_{B\pm} \label{DB}%
\end{align}
where the meaning of $g_{\pm}$ is $g_{\pm}=g_{x}\pm ig_{y}$ and similarly for
$j_{E\pm}$ and $j_{B\pm}$. Recall that there are two sets of Eqs.
(\ref{Dvz})-(\ref{DB}), one set for each particle species. These equations
imply the following evolution equation for $E_{\pm}$%
\begin{equation}
\hat{D}_{E}E_{\pm}=S_{E} \label{DE}%
\end{equation}
where the wave propagator is
\begin{align}
\hat{D}_{E}  &  \equiv\hat{G}_{e}\hat{G}_{i}\square+\omega_{pi}^{2}\hat{G}%
_{e}\hat{d}_{i}+\omega_{pe}^{2}\hat{G}_{i}\hat{d}_{e}\label{Dop_E}\\
\square &  \equiv\partial_{t}^{2}-\partial_{z}^{2}\nonumber\\
\hat{d}_{(i)}  &  \equiv\partial_{t}+v_{z(i)}\partial_{z}\nonumber\\
\hat{G}_{(i)}  &  \equiv\partial_{t}+v_{z(i)}\partial_{z}\pm i\omega
_{c(i)}\nonumber
\end{align}
and the source term is%
\begin{align}
S_{E}  &  \equiv-q_{i}n\hat{G}_{e}\partial_{t}g_{i\pm}-q_{e}n\hat{G}%
_{i}\partial_{t}g_{e\pm}\nonumber\\
&  \pm i\left(  \omega_{pi}^{2}v_{iz}\hat{G}_{e}+\omega_{pe}^{2}v_{ez}\hat
{G}_{i}-\hat{G}_{e}\hat{G}_{i}\partial_{z}\right)  j_{B\pm}\nonumber\\
&  -\hat{G}_{e}\hat{G}_{i}\partial_{t}j_{E\pm} \label{S}%
\end{align}
In the absence of the gravitational source terms the evolution equation
(\ref{DE}) reduces to $\hat{D}_{E}E_{\pm}=0$ and for this case we in
particular note the solution \cite{Stenflo}
\begin{align}
E_{\pm}  &  =Ee^{i(kz-\omega t)}\label{EMsol1}\\
B_{\pm}  &  =\pm i\frac{k}{\omega}E_{\pm}\label{EMsol2}\\
v_{\pm}  &  =i\Lambda E_{\pm}\label{EMsol3}\\
\Lambda &  \equiv\frac{q}{\gamma m}\frac{\omega-kv_{z}}{\omega(\omega
-kv_{z}\mp\omega_{c})}\label{Lambda}\\
\gamma &  =(1-v_{\pm}v_{\mp}-v_{z}^{2})^{-\frac{1}{2}} \label{gamma}%
\end{align}
with the dispersion relation
\begin{align}
0  &  =D_{E}\equiv\left(  \omega-kv_{ze}\mp\omega_{ce}\right)  \left(
\omega-kv_{zi}\mp\omega_{ci}\right)  \left(  \omega^{2}-k^{2}\right)
\nonumber\\
&  -\omega_{pi}^{2}\left(  \omega-kv_{ze}\mp\omega_{ce}\right)  \left(
\omega-kv_{zi}\right) \nonumber\\
&  -\omega_{pe}^{2}\left(  \omega-kv_{zi}\mp\omega_{ci}\right)  \left(
\omega-kv_{ze}\right)  \label{DRElong}%
\end{align}
that can be written more compact as
\begin{equation}
\omega^{2}-k^{2}=\sum_{(i)}\frac{\omega_{p(i)}^{2}\left(  \omega
-kv_{z(i)}\right)  }{\omega-kv_{z(i)}\mp\omega_{c(i)}} \label{DRE}%
\end{equation}
We again emphasize that the solution is valid for arbitrarily large
amplitudes. The condition for the plasma to be electrically neutral, $j^{0}%
=0$, implies $\gamma_{e}n_{e}=\gamma_{i}n_{i}$. As $\gamma_{(i)}n_{(i)}$ is a
preserved quantity (by Eq. (\ref{plasma5})), the proper density, $n_{(i)}$, is
not. Accordingly, both $\omega_{p(i)}$ and $\omega_{c(i)}$ depend on
$\gamma_{(i)}$. This makes the dispersion relation amplitude dependent.

For future reference we devote the remainder of this section to establish some
properties of the free {\small EMW}s. As we will focus on {\small EMW}s with
$\omega=k$ we note that the dispersion relation in this case implies the
following relation between the background magnetic field and the drift
velocities
\begin{equation}
\Omega_{ce}=\pm(\gamma_{e}+\frac{m_{i}}{m_{e}}\gamma_{i})\frac{(1-v_{ze}%
)(1-v_{zi})}{v_{ze}-v_{zi}}\label{B-relation}%
\end{equation}
where $\Omega_{ce}\equiv\gamma_{e}\omega_{ce}/\omega=\omega^{-1}q_{e}%
B_{(0)}/m_{e}$ is the normalized (and preserved) ``electron'' cyclotron
frequency and we have made use of the neutrality condition $\gamma_{e}%
n_{e}=\gamma_{i}n_{i}$. Note that for any given background plasma parameters
(except the absence of any drifts), there always exist a wave frequency
fulfilling Eq. (\ref{B-relation}). The components of the perturbed
energy-momentum tensor, for this large amplitude circularly polarized
{\small EMW}, that couples to {\small GW}s has the values
\begin{align}
\delta T_{11} &  =-{\tfrac{1}{2}}\left(  E_{x}^{2}+B_{x}^{2}-E_{y}^{2}%
-B_{y}^{2}\right)  +\sum\gamma mnv_{x}^{2}\label{dT11_E}\\
\delta T_{22} &  ={\tfrac{1}{2}}\left(  E_{x}^{2}+B_{x}^{2}-E_{y}^{2}%
-B_{y}^{2}\right)  +\sum\gamma mnv_{y}^{2}\label{dT22_E}\\
\delta T_{12} &  =-\left(  E_{x}E_{y}+B_{x}B_{y}\right)  +\sum\gamma
mnv_{x}v_{y}\label{dT12_E}%
\end{align}
Because the wave is circularly polarized and transversal it is effectively of
linear nature (the Lorentz force is linear in the wave amplitude) and has no
density (nor pressure) perturbations. Therefore, the wave energy density can
be calculated by means of Eq. (\ref{EEM}). For the given wave perturbations,
the wave energy density is given by%
\begin{align}
\mathcal{E}_{E} &  =\frac{1}{2\omega}\partial_{\omega}(\omega^{2}A_{\pm
})|E_{\pm}|^{2}\label{E(A)}\\
A_{\pm} &  \equiv1-\sum\frac{\omega_{p}^{2}}{\omega^{2}}\frac{\omega-kv_{z}%
}{\omega-kv_{z}\mp\omega_{c}}\nonumber
\end{align}
Applying the dispersion relation (\ref{DRE}) with $\omega=k$, the wave energy
density can, after some algebra, be formulated as%
\begin{align}
\mathcal{E}_{E} &  =\frac{1}{2\omega^{2}\alpha_{e}\alpha_{i}}\left[
2\omega^{2}\alpha_{e}\alpha_{i}-\omega_{pe}^{2}(1-v_{e}+\alpha_{i})\right.
\nonumber\\
&  \left.  -\omega_{pi}^{2}(1-v_{i}+\alpha_{e})\right]  \left|  E_{\pm
}\right|  ^{2}\label{EE}%
\end{align}
where we have introduced the abbreviation
\begin{equation}
\alpha_{(i)}\equiv1-v_{(i)}\mp\omega_{c(i)}/\omega
\end{equation}

\begin{center}
\includegraphics[
height=2.5642in,
width=3.237in
]%
{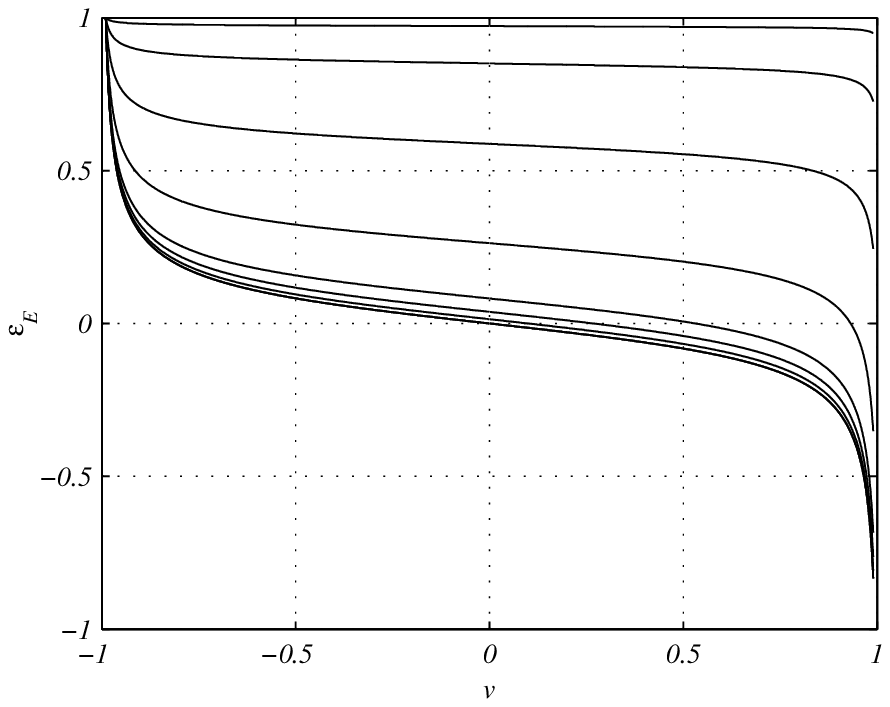}%
\\
FIG.1. \emph{The dependence of wave energy $\mathcal{E}_{E}$ (normalized by $\left|
E_{\pm}\right|  ^{2}$ and rescaled) on drift flow $v\equiv v_{i}=-v_{e}$ for
null geodesic small amplitude waves in an electron-ion plasma ($m_{e}%
/m_{i}\sim1000$). The different curves, from the top and down, corresponds to
the plasma frequencies $\Omega_{pe}=2,5,10,20,40,60,100,1000$. As the sign of
the energy is the main focus here, the energy in the different cases are
rescaled by suitable factors.}%
\end{center}

\vspace{0.5cm}

It is important to note here that $\mathcal{E}_{E}$ is, in contrast to
$\mathcal{E}_{GW}$ and $\mathcal{E}_{X}$, not positive definite. For some
background parameters the wave energy density may be negative \cite{Weiland}.
This is due to the presence of the velocity drifts, providing the system with
free energy. Negative energy waves are wave perturbations such that the total
energy of the plasma is less than the energy of the unperturbed state
\cite{Strictly}. To show that there are waves fulfilling $\omega=k$ and
$\mathcal{E}_{E}<0$ simultaneously, we consider an electron-ion plasma
($m_{e}/m_{i}\sim1000$) with drift flows $v_{e}=-v_{i}$. In FIG 1 the wave
energy factor $\mathcal{E}_{E}/\left|  E_{\pm}\right|  ^{2}$ (for small
amplitude waves) is presented, as a function of ion drift velocity, for a
number of different values of the normalized ``electron'' plasma frequency
$\Omega_{pe}\equiv\gamma_{e}\omega_{pe}/\omega$ and with the cyclotron
frequency fulfilling the condition (\ref{B-relation}). As the sign of the
energy is the main focus here, the energy factor in the different cases are
rescaled by suitable factors. Negative wave energy occurs for positive ion
velocities \cite{Velocity}. The minimum required ion velocity depends
inversely on the plasma frequency. For $\Omega_{pe}=1000$, negative wave
energy requires only $v_{i}\gtrsim10^{-4}$ whereas for $\Omega_{pe}\lesssim20$
it requires highly relativistic velocities, $v_{i}\gtrsim0.95$.

\section{Resonant wave interaction\label{Res wave int}}

The wave evolution equations (\ref{hA}), (\ref{hB}), (\ref{Xwaveq}) and
(\ref{DE}) where derived with no assumptions on the specific wave forms. In
this section we assume the wave frequencies and the (uniform) background
plasma to be in a state where wave resonance occurs, since this gives the most
efficient energy transfer, and we derive the corresponding wave interaction
equations. We let
\begin{align*}
h_{A,B} &  =\tilde{h}_{A,B}+c.c.\\
\delta B_{x} &  =\tilde{B}+c.c.\\
E_{\pm} &  =E_{x}\pm iE_{y}%
\end{align*}
with $\tilde{h}_{A,B}=\frak{h}_{A,B}e^{i\left.  k_{G}\right.  _{a}x^{a}}$,
$\tilde{B}=\frak{B}e^{i\left.  k_{X}\right.  _{a}x^{a}}$ and $E_{\pm}%
=\frak{E}e^{i\left.  k_{E}\right.  _{a}x^{a}}$ and the four-wave vectors
$k_{G}^{a}=(\omega_{G},0,0,k_{G})$, $k_{X}^{a}=(\omega_{X},0,0,k_{X})$, and
$k_{E}^{a}(\omega_{E},0,0,k_{E})$ are assumed to fulfill the dispersion
relations $\left.  k_{G}\right.  _{a}k_{G}^{a}=0$, Eq. (\ref{DX}) and Eq.
(\ref{DRE}), respectively. As before, $c.c.$ stands for the complex conjugate
of the preceding term. The conditions for the {\small EMW}s to be resonant
with a {\small GW} are, in the case of perpendicular propagation
\begin{equation}
k_{G}^{a}=k_{X}^{a}%
\end{equation}
and in the case of parallel propagation
\begin{equation}
k_{G}^{a}=2k_{E}^{a}\label{matchingGWEMW}%
\end{equation}
Here we have used that the extraordinary {\small EMW} contributes linearly to
the perturbed energy momentum tensor (\ref{dT11_B}) and the circular polarized
large amplitude {\small EMW} quadratically in (\ref{dT11_E})-(\ref{dT12_E})
and thus produces terms proportional to $e^{i2(k_{E}z-\omega_{E}t)}$.
Consistently, the source terms for the extraordinary {\small EMW} is linear
(in the {\small GW} amplitude) whereas for the circular polarized large
amplitude {\small EMW} there are terms proportional to $E_{\pm}h_{A,B}$. Since
(in the high frequency limit) $\omega_{G}/k_{G}=1$, also the {\small EMW}s
should have $\omega_{X}/k_{X}=1$ and $\omega_{E}/k_{E}=1$ in order to remain
in phase with the {\small GW}. The interaction is weak in the sense that the
coupling terms in the evolution equations (\ref{hA}), (\ref{hB}),
(\ref{Xwaveq}) and (\ref{DE}) are either proportional to the small {\small GW}
amplitude or the gravitational coupling constant. This leads to a weak time
dependence in the wave amplitudes $\frak{h}$, $\frak{B}$ and $\frak{E}$, i.e.
$\partial_{t}\frak{h}\ll\omega_{G}\frak{h}$ etc. We will use the notation of
``slow derivatives'', $\tilde{\partial}_{t}$, that acts only on the wave
amplitude, so that e.g. $\partial_{t}h_{A}=i\omega_{G}h_{A}+\tilde{\partial
}_{t}h_{A}$, and we will neglect slow derivatives that are of second or higher
order. Consistent with this level of approximation, we make use of the
relations between the perturbations valid for free waves, i.e. Eqs
(\ref{EMsol1})-(\ref{EMsol3}), in the computation of the source terms.
Corrections to this approximations produces higher order source terms.

We now treat the two cases of propagation direction separately.

\subsection{Perpendicular propagation}

From the dispersion relation (\ref{DX}) it follows that in order for the
extraordinary {\small EMW} to propagate along null geodesics the wave
frequencies must be equal to the plasma frequency, $\omega\equiv\omega_{pe}$.
Although this is a special case, a {\small GW} or an {\small EMW} propagating
in a slowly varying background density may often reach such resonance regions.
Assuming resonant waves, we take $k_{G}^{a}=k_{X}^{a}=(\omega,0,0,\omega)$.
The wave evolution equations (\ref{hA}) and (\ref{Xwaveq}) then imply the
following wave interaction equations (dropping the tilde notation on the waves
and denoting $h\equiv h_{A}$ and $B\equiv\delta B$)
\begin{align}
\tilde{\partial}_{t}h  &  =-iC_{G}B\label{wieqX1}\\
\tilde{\partial}_{t}B  &  =-iC_{X}h \label{wieqX2}%
\end{align}
where $C_{G}\equiv\omega^{-1}\kappa B_{(0)}$ and $C_{X}\equiv\tfrac{1}%
{2}\omega\omega_{h}^{-2}\omega_{c}^{2}B_{(0)}$. From Eqs. (\ref{EGW}) and
(\ref{EX}) the total wave energy is $\mathcal{E}_{GW}+\mathcal{E}_{X}%
=\omega^{2}\kappa^{-1}|h|^{2}+2\omega_{c}^{-2}\omega_{h}^{2}\left|  B\right|
^{2}$ and it follows from Eq. (\ref{wieqX1}) and (\ref{wieqX2}) that this is a
conserved quantity. The interaction equations has the following solution for
the wave amplitudes
\begin{align}
B  &  =B_{0}\cos(\psi t)+iC_{X}\psi^{-1}h_{0}\sin(\psi t)\label{solX1}\\
h  &  =h_{0}\cos(\psi t)+iC_{G}\psi^{-1}B_{0}\sin(\psi t) \label{solX2}%
\end{align}
where $B_{0}\equiv B(t=0)$ and $h_{0}\equiv h(t=0)$, $\psi\equiv\sqrt
{C_{X}C_{G}}=\sqrt{\frac{\kappa}{2}}|\omega_{h}^{-1}\omega_{c}B_{(0)}|$. The
solution shows that the total wave energy alternates between the {\small GW}
and the {\small EMW} with the frequency $\psi$. In most applications $\psi$ is
very small and it is then meaningful to linearize the trigonometric functions
in $\psi$ and it is clear that, for time scales smaller than $\psi^{-1}$,
{\small GW}s are converted into {\small EMW}s as $h=C_{G}B_{0}t$ or,
alternatively, {\small EMW}s are converted into {\small GW}s as $B=C_{X}%
h_{0}t$. It should be noted that in the later case the time $t_{nl}$ for the
{\small EMW} perturbation to reach the nonlinear stage, $B\sim B_{(0)}$, is
typically much smaller than total energy conversion time, $\psi^{-1}$.

\subsection{Parallel propagation}

For resonant interaction between {\small EMW}s and {\small GW}s in the
parallel case we take $k_{G}^{a}=2k_{E}^{a}=(2\omega,0,0,2\omega)$. As the
{\small EMW}s that we are considering are circularly polarized waves, we also
introduce variables for circularly polarized {\small GW}s $h_{\pm}\equiv
h_{A}\pm ih_{B}$. If the {\small GW} and the {\small EMW} are oppositely
polarized the coupling vanishes, so from here on we assume identical
polarization. The evolution Eqs. (\ref{hA}) and (\ref{hB}) combines to
\begin{equation}
\left(  -\partial_{t}^{2}+\partial_{z}^{2}\right)  h_{\pm}=-\sum\kappa
mn\gamma^{2}v_{\pm}^{2}%
\end{equation}
and together with the {\small EMW} evolution equation (\ref{DE}) this implies
the following interaction equations for the variables $h\equiv h_{\pm}$ and
$E\equiv E_{\pm}$%
\begin{align}
\tilde{\partial}_{t}h  &  =iC_{{\small GW}}E^{2}\label{wieqE1}\\
\tilde{\partial}_{t}E  &  =iC_{E}E^{\ast}h \label{wieqE2}%
\end{align}
where
\begin{align}
C_{{\small GW}}  &  \equiv\frac{\kappa}{4\omega}\sum mn\gamma^{2}\Lambda^{2}\\
C_{E}  &  \equiv\omega^{2}\left[  \gamma_{i}\frac{m_{i}}{q_{i}}\Lambda
_{i}\omega_{pi}^{2}(1-v_{zi})\alpha_{e}\right. \nonumber\\
&  \left.  +\gamma_{e}\frac{m_{e}}{q_{e}}\Lambda_{e}\omega_{pe}^{2}%
(1-v_{ze})\alpha_{i}\right] \nonumber\\
&  \times\left[  2\omega^{2}\alpha_{e}\alpha_{i}-\omega_{pi}^{2}\left(
1-v_{zi}+\alpha_{e}\right)  \right. \nonumber\\
&  \left.  -\omega_{pe}^{2}\left(  1-v_{ze}+\alpha_{i}\right)  \right]  ^{-1}%
\end{align}
In the derivation we have noted that the effective currents, $\mathbf{j}_{E}$
and $\mathbf{j}_{B}$, vanishes and used that $g_{\pm}=-{\tfrac{1}{2}}%
\gamma(1-v_{z})v_{\pm}^{\ast}\partial_{t}h_{\pm}$. The coupling coefficients
can be related to the wave energy densities, Eqs. (\ref{EGW}) and (\ref{EE}),
according to%
\begin{align}
C_{{\small GW}}  &  =\mathcal{CE}_{{\small GW}}^{-1}\left|  h\right|  ^{2}\\
C_{E}  &  =\mathcal{CE}_{E}^{-1}\left|  E\right|  ^{2}%
\end{align}
where
\begin{equation}
\mathcal{C\equiv}\frac{1}{2\omega}\sum\frac{\omega_{p(i)}^{2}}{\alpha_{(i)}%
^{2}}\left(  1-v_{z(i)}\right)  ^{2}%
\end{equation}
It is then straightforward to confirm that Eqs. (\ref{wieqE1}) and
(\ref{wieqE2}) implies that the total wave energy density
\begin{align}
\mathcal{E}  &  =\frac{2\omega^{2}}{\kappa}|h|^{2}+\frac{1}{2\omega^{2}%
\alpha_{e}\alpha_{i}}\left[  2\omega^{2}\alpha_{e}\alpha_{i}\right.
\nonumber\\
&  \left.  -\omega_{pi}^{2}\left(  1-v_{zi}+\alpha_{e}\right)  -\omega
_{pe}^{2}\left(  1-v_{ze}+\alpha_{i}\right)  \right]  |E|^{2}%
\end{align}
is conserved.

The solution to the wave interaction equations (\ref{wieqE1}) and
(\ref{wieqE2}) can be given in terms of Jacobi elliptic functions. We exploit
the fact that the system can be reformulated as a three-wave interacting
system (see e.g. \cite{Weiland}) with two identical {\small EMW}s and one
{\small GW}. This is done by taking $E_{2}=E_{3}\equiv\frac{1}{2}E$ such that
Eq. (\ref{wieqE2}) can be splitted into two (identical, but differently
labeled) equations and the right hand side of Eq. (\ref{wieqE1}) reads
$i4C_{GW}E_{2}E_{3}$. Next we convert the obtained three-wave interaction
equations to a real system and renormalize the amplitudes such that the
coupling coefficients become unity. The real system is \cite{Phase} (dropping
the tilde notation)
\begin{align}
\partial_{t}\psi_{1}  &  =s_{1}c_{1}\psi_{2}\psi_{3}\cos(\phi)\label{real1}\\
\partial_{t}\psi_{2}  &  =s_{2}c_{2}\psi_{1}\psi_{3}\cos(\phi)\label{real2}\\
\partial_{t}\psi_{3}  &  =s_{3}c_{3}\psi_{1}\psi_{2}\cos(\phi) \label{real3}%
\end{align}
where we have taken $ih=\psi_{1}e^{i\phi_{1}}$, $E_{n}=\psi_{n}e^{i\phi_{n}}$,
$-4C_{GW}=s_{1}c_{1}$, $C_{E}=s_{n}c_{n}$ (for $n=2,3$) with $\psi_{1}=|h|$,
$\psi_{n}=|E_{n}|$, $c_{1}=|C_{GW}|$, $c_{n}=|C_{E}|$, $\phi=\phi_{1}-\phi
_{2}-\phi_{3}$ and, $s_{1}$, $s_{2}$ and $s_{3}$ are the signs of $-C_{GW}$
and $C_{E}$, respectively. The coupling coefficients $c_{1}$, $c_{2}$ and
$c_{3}$ are made unity by the renormalization
\begin{align}
\psi_{1}  &  \rightarrow\Psi_{1}\equiv\sqrt{c_{2}c_{3}}\psi_{1}
\label{renorm1}\\
\psi_{2}  &  \rightarrow\Psi_{2}\equiv\sqrt{c_{3}c_{1}}\psi_{2}
\label{renorm2}\\
\psi_{3}  &  \rightarrow\Psi_{3}\equiv\sqrt{c_{1}c_{2}}\psi_{3}%
\end{align}
and we also note that we may, with no loss in generality, assume $s_{2}%
=s_{3}=1$ (it is only the sign of $s_{1}s_{2}=s_{1}s_{3}$ that matters because
an overall sign can be removed by a renormalization). The general solution can
then be found in litterature, e.g. in Ref. \cite{Weiland}. The strongest
coupling occurs for $\cos\phi=\pm1$ and in this case there are some
particularly simple solutions that are listed below (recall that $\Psi
_{2}=\Psi_{3}$).

\begin{enumerate}
\item [\emph{a)}]{\small EMW} to {\small GW} conversion with positive
{\small EMW} energy\newline
\[
s_{1}=-1,\;\Psi_{1}(0)=0,\;\Psi_{2,3}(0)=\Psi\;\text{and\ }\cos\phi=-1
\]%
\begin{equation}
\Psi_{1}=\Psi\text{tanh}(\Psi t)\;\;\;\;\;\;\;\Psi_{2,3}=\Psi\text{sech}(\Psi
t) \label{Psi_a}%
\end{equation}

\item[\emph{b)}] {\small EMW}-{\small GW} instability with no initial
{\small GW}\newline
\[
s_{1}=1,\;\Psi_{1}(0)=0,\;\Psi_{2,3}(0)=\Psi\;\text{and\ }\cos\phi=1
\]%
\begin{equation}
\Psi_{1}=\Psi\text{tan}(\Psi t)\;\;\;\;\;\;\;\Psi_{2,3}=\Psi\text{sec}(\Psi t)
\label{Psi_b}%
\end{equation}

\item[\emph{c)}] {\small EMW}-{\small GW} instability with equal initial
amplitudes\newline
\[
s_{1}=1,\;\Psi_{1}(0)=\Psi_{2,3}(0)=\Psi\;\text{and}\;\cos\phi=1
\]%
\begin{equation}
\Psi_{1}=\left(  \Psi^{-1}-t\right)  ^{-1}\;\;\;\;\;\;\;\Psi_{2,3}=\left(
\Psi^{-1}-t\right)  ^{-1} \label{Psi_c}%
\end{equation}

\item[\emph{d)}] {\small EMW}-{\small GW} interaction with decaying
amplitudes\newline
\[
s_{1}=1,\;\Psi_{1}(0)=\Psi_{2,3}(0)=\Psi\;\text{and\ }\cos\phi=-1
\]%
\begin{equation}
\Psi_{1}=\left(  \Psi^{-1}+t\right)  ^{-1}\;\;\;\;\;\;\;\Psi_{2,3}=\left(
\Psi^{-1}+t\right)  ^{-1} \label{Psi_d}%
\end{equation}
\end{enumerate}

Also conversion from {\small GW} to {\small EMW} energy occurs (given that
there is a small initial {\small EMW} perturbation) the solution is somewhat
more complicated than the ones listed above \cite{Weiland}. Note that
$\Psi_{1}(0)=\Psi$ and $\Psi_{2,3}(0)=0$ only has the trivial solution
$\Psi_{1}=\Psi$ and $\Psi_{2,3}=0$. The solution \emph{a)} corresponds to an
{\small EMW} being converted into a {\small GW} on the characteristic time
scale $\pi/2\Psi$. The solutions \emph{b)} and \emph{c)} are explosive
solutions, i.e. the wave amplitudes diverges on a finite time $t=\pi/2\Psi$
and $t=1/\Psi$, respectively. These explosive instabilities can only occur for
$s_{1}=s_{2}=s_{3}$, which is equivalent to saying that the {\small GW} and
{\small EMW} energy must be of different sign. The explosive instability thus
relies on the existence of negative energy {\small EMW}s and it is the free
energy connected with the background drift flow that feeds the instability. We
return to these interesting cases in section \ref{Summary disc}. The solution
\emph{d)} also involves a negative energy {\small EMW}, but in this case both
the {\small EMW} and the {\small GW} amplitudes tends to zero as $(\Psi
^{-1}+t)^{-1}$. This means that wave energy is converted into free energy of
the background state, i.e. acceleration of the drift flow. As for the cases
when the wave phases are not fulfilling $\cos\phi=\pm1$ the behaviour is
similar \cite{Phase} but with somewhat larger characteristic times.

We now explore the possibility, indicated by the solution \emph{d)}, that the
waves may interact to accelerate the background drift flow. For the wave
perturbations we are considering here, the fluid equation of motion
(\ref{plasma6}) has the property $v_{i}\partial_{t}\gamma v^{i}=g_{z}$ (where
$i=1,2,3$). This implies that $\xi\equiv\gamma(1-v_{z})$ is an \emph{exactly}
conserved quantity, as is seen by making the computation
\begin{equation}
\partial_{t}\xi\equiv\partial_{t}\gamma-\partial_{t}\gamma v_{z}=v_{i}%
\partial_{t}\gamma v^{i}-g_{z}=0 \label{conservation}%
\end{equation}
Making use of this, the $z$-component of the fluid momentum equation (Eq.
(\ref{plasma6})) may be rewritten as
\begin{align}
\partial_{t}v_{z}  &  ={\tfrac{i\omega}{2}}\xi\Lambda^{2}\left(  E^{2}h^{\ast
}-E^{\ast2}h\right) \nonumber\\
&  =-{\omega}\xi\Lambda^{2}|E|^{2}|h|\cos(\phi) \label{driftacc}%
\end{align}
where the phase angle $\phi$ has the same meaning as in Eqs. (\ref{real1}%
)-(\ref{real3}). From this equation it is clear that the drift flow may be
accelerated or decellerated in the $z$-direction by the interacting waves.
Maximum acceleration occurs for $\cos\phi=-1$. In the \emph{pump-wave
approximation}, where the waves are assumed energetic and can be taken to be
of constant amplitude (an approximation breaking down when the energy of the
drift flow becomes comparable with the wave energies), the drift flow $v_{z}$
grows linearly in time as (for $\cos\phi=-1$)
\begin{equation}
v_{z}=v_{z}(0)+{\omega}\xi\Lambda^{2}|E|^{2}|h|t
\end{equation}
and we point out that for relativistic {\small EMW} amplitudes $\Lambda
^{2}|E|^{2}=|v_{\pm}|^{2}\approx1$.

\section{\label{Summary disc}Summary and discussion}

In this paper we have studied the interaction between {\small GW}s and
{\small EMW}s in magnetized plasmas for perpendicular and parallel propagtion
with respect to the background magnetic field. The wave evolution equations
were derived in section \ref{GW evol} and \ref{EM evol}. An important point
here which, as far as we know, has not been fully recognized previously is
that, given the high frequency approximation, the {\small GW}s can be taken to
be in the TT gauge even in the presence of matter. Consequently, the deduction
of the {\small GW} source term associated with the {\small EMW} perturbed
energy-momentum tensor is simplified, see Eqs. (\ref{hA}) and (\ref{hB}). The
{\small GW}, on the other hand, produces effective currents in the Maxwell
equations and an effective gravitational force on the plasma, resulting in a
source term in the {\small EMW} evolution equations (\ref{Xwaveq}) and
(\ref{DE}). Furthermore, it was shown in section \ref{EM evol} that, provided
there is a velocity drift in the background state, the wave energy may be
negative \cite{Weiland} for {\small EMW}s with phase velocity equal to $c=1$
(when propagating parallel to the background magnetic field).

Resonant interaction (involving a single {\small EMW}) may occur when the
frequencies matches and the phase velocity of the {\small EMW} coincide with
that of the {\small GW}, which to a very good precision equals $c=1$ in the
high-frequency approximation. The wave interaction equations governing the
resonant interaction process were derived and analyzed in section \ref{Res
wave int} for the cases of perpendicular (high-frequency extraordinary
{\small EMW}s) and parallel propagation (finite amplitude circular polarized
{\small EMW}s) with respect to the background magnetic field. In both cases
the interaction equations were shown to imply conservation of total wave energy.

\emph{i) Perpendicular propagation.} The wave interaction equations for this
case are Eqs. (\ref{wieqX1}) and (\ref{wieqX2}). The solution, given by Eqs.
(\ref{solX1}) and (\ref{solX2}), reveals that the conversion rate
\cite{Conversion rate} is of order $\psi=(\kappa/2)^{1/2}\omega_{c}\omega
_{h}^{-1}B_{(0)}$. The effect of the plasma is to diminish the coupling
strength. The conversion rate becomes $\psi=(\kappa/2)^{1/2}B_{(0)}$ for
strong magnetic fields, i.e. $\omega_{c}\gg\omega_{p}$ (similar to having no
plasma present), whereas for weak magnetic fields, $\omega_{c}\ll\omega_{p}$,
the conversion rate is a factor $\omega_{c}/\omega_{p}$ smaller. Although
$\psi$ is typically small, {\small GW}s may still generate {\small EMW}s with
significant amplitude. For small $\psi t$, the {\small EMW} are converted into
{\small GW}s as $B=\frac{1}{2}\omega_{p}\omega_{c}^{2}\omega_{h}^{-2}%
B_{(0)}h_{0}t$.

\emph{ii) Parallel propagation with positive {\small EMW} energy.} The
solutions of the wave interaction equations (\ref{wieqE1}) and (\ref{wieqE2})
were found by reformulating them as a three-wave interacting system (with two
identical {\small EMW}s) for which the general solutions are well known. From
the solution (\ref{Psi_a}) it follows that the conversion rate
\cite{Conversion rate} for conversion from {\small EMW} to {\small GW} is
$\Psi=\sqrt{C_{GW}C_{E}}|E_{ini}|$, where $E_{ini}$ is the initial electric
field amplitude. The conversion rate increases with increasing initial
amplitude and with increasing drift velocity (through $\gamma$). In order to
have more transparent formulas we consider a special case, namely
low-frequency waves, such that $\omega\ll\omega_{ci},\omega_{ce}$. The
conversion rate can be formulated as%
\begin{equation}
\Psi=\left(  C_{GW}C_{E}^{2}\mathcal{C}^{-1}\mathcal{E}_{E,ini}\right)
^{\frac{1}{2}}\label{lfPsi}
\end{equation}
and in the case of low-frequency waves%
\begin{align*}
C_{GW} &  =\frac{\kappa}{2\omega^{2}}\mathcal{C}\\
C_{E} &  =\mathcal{CD}^{-1}\\
\mathcal{E}_{E} &  =\mathcal{D}\left|  E_{\pm}\right|  ^{2}%
\end{align*}
where%
\[
\mathcal{C}=\frac{\omega}{2}\sum\frac{\omega_{p}^{2}}{\omega_{c}^{2}}%
(1-v_{z})^{2}\;\;,\;\;\;\;\mathcal{D}=1\pm\frac{1}{2\omega}\sum\frac
{\omega_{p}^{2}}{\omega_{c}}%
\]
and the final {\small GW} amplitude is given by $h_{final}=(\mathcal{E}%
_{E,ini}\kappa/2\omega^{2})^{1/2}$.

\emph{iii) Parallel propagation with negative {\small EMW} energy.} As was
remarked previously, the {\small EMW} energy density (given by Eq. (\ref{EE}))
may be negative. Such waves are perturbations with the property that the
energy of the perturbed system is less than for the unperturbed system
\cite{Strictly}. As is clear from FIG 1, this occurs most commonly for
low-frequency waves in electron-ion plasmas, i.e. $\omega\ll\gamma_{e}%
\omega_{pe}$. The wave energy conservation then allows \emph{simultaneous}
growth of the {\small GW} and {\small EMW} amplitudes. The physical
interpretation is that the background flow is unstable with respect to this
type of perturbations and the energy associated with the flow is converted
into wave energy. Given that there is an ion flow in the direction of wave
propagation and a large enough plasma frequency, this instability will always
produce waves with a unique EMW frequency determined by Eq. (\ref{B-relation})
and GW frequency given by the matching condition (\ref{matchingGWEMW}). It
should be noted that this is a \emph{nonlinear} instability and would not have
been found using conventional linear stability analysis. The solutions
(\ref{Psi_b}) and (\ref{Psi_c}) shows that the background is
\emph{explosively} unstable with respect to {\small GW}-{\small EMW}
perturbations, the amplitudes reaches infinity in a finite time $t\sim
\Psi^{-1}$ (the interaction equations are of course invalidated before this
time is reached). The solution (\ref{Psi_d}) is an example of the inverse
process, where both waves decreases in amplitude. This means that wave energy
is converted into free energy of the background state, i.e. the drift flow is
accelerated. An evolution equation for the background flow was also derived,
see Eq. (\ref{driftacc}). The response of the plasma on the interacting waves
is an acceleration (if $\cos\phi>0$) or deceleration (if $\cos\phi<0$). The
conclusion is that {\small GW}s and {\small EMW}s can interact to produce a
drift flow along the background magnetic field with initial linear growth rate
as large as $\omega\gamma(1-v_{z})|h|$. The magnitude of the produced drift
flow can be estimated from the initial wave energies and the time scale for
all the wave energy to be converted into drift flow is at least of the order
given by Eq. (\ref{Psi_d}), $t\sim\Psi^{-1}$.

Finally we discuss the relevance of the results presented in this paper to
various astrophysical and cosmological processes. The phenomena studied in
this paper involve energy transfer between {\small GW}s, {\small EMW}s and a
background flow. This could in principle be an important feature in any
scenario involving magnetized plasmas, energetic {\small GW}s and
{\small EMW}s and/or strong background plasma flows, e.g. supernova
explosions, gamma-ray bursts, jets of accreting condensed objects, phenomena
in the vicinity of neutron stars and in the primordial fluctuations of the
cosmological plasma.
%%%%%%%%
% Added text, numerical estimation
The energy conversion must take
place within reasonable times and/or volumes for the considered processes to
be of significance. As a specific numerical example let us first consider
graviational to electromagnetic conversion for \emph{perpendicular} propagation. For
the case of {\small GW}s from a binary pulsar close to collapse (corresponding
to a wavelength of, let us say, $\lambda\sim10^{5}\mathrm{m}$) we may have $h\sim0.001$ a few
wave-lengths away from the pulsar. A rough estimation based on Eq. (\ref{solX1}) then
gives a distance $L\sim\lambda/h\sim10^{8}\mathrm{m}$ for the induced magnetic
field to be comparable to the unperturbed field \cite{BVPnote}. A significant transfer of {\small GW}-energy is not realistic
in this example, however, since the background energy density is too low. As
another example, let us consider {\small EMW }to {\small GW }conversion for
\emph{parallell} propagation and positive wave energy in the low-frequency regime.
Let us take $B_{0}\sim10^{-3}\mathrm{T}$ and $n_{0}\sim10^{18}\mathrm{m}^{-3}$.
Naturally, the beam velocities are assumed to fulfill (\ref{B-relation}), and the wave
frequency is assumed to obey $\omega\ll$  $\omega_{ci}\sim2\times
10^{5}\mathrm{s}^{-1}$. Using Eq. (\ref{lfPsi}) the characteristic time scale for energy
conversion then becomes $t\sim\Psi^{-1}\sim10^{7}\left(  \mathcal{E}%
_{E,ini}\right)  ^{-1/2}\mathrm{s}$, where the electromagnetic wave energy density should be
given in SI-units. Apparently we need wave energy densities $\mathcal{E}%
_{E,ini}\sim10^{14}\mathrm{J/m}^{3}$ to get conversion within $1$ second.
Note that {\small EMW} energy densities in the laboratory can be magnitudes larger than
this \cite{Morou} and the value is also modest as compared to some astrophysical events,
like gamma-ray bursts. However, we note that the required wave energy content
is still very large, since the {\small EMW} must be distibuted in a volume not less
than a cube with a side of 1 light second, for efficient {\small GW} conversion to take
place.
%%%%%%%%

The model we have presented is idealized in several ways,
e.g. we have assumed a uniform background and monochromatic waves. The model
can be extended to include weakly nonuniform backgrounds, where the
interaction can be treated using mode conversion theory, see e.g. Ref.
\cite{Mode conv.} and Refs. therein\textbf{. }It should also be noted that the
background drift flow is in general also unstable with respect to other types
of perturbations, e.g. ion-accoustic perturbations which typically may have a
higher growth rate. With a more realistic velocity distribution among the
particles, including a thermal spread, the conditions and growth rates for
various instabilities change. It is an open question whether {\small GW}%
-{\small EMW} instabilities (of the type presented in this paper) can be the
only ones (or the dominating ones) in certain regions of parameter space. We
have also treated the gamma factors and the background flow as constants in
time. This is not a valid approximation for the long term evolution when the
change in energy becomes comparable to the initial energy of the source. An
exception to this case is interaction between waves with ``moderate'' positive
energies in the presence of a highly energetic background flow, such that
$\gamma\approx(1-v_{z}^{2})^{-\frac{1}{2}}$ and $v_{z}$ are approximately
constant for all times. Variations in $\gamma$ ad $v_{z}$ can, as these occurs
in the dispersion relation (\ref{DRE}), also affect the frequency of the
{\small EMW}, thereby invalidate the assumed perfect resonance and thus
reducing the coupling strength.

In conclusion, energy transfer between {\small GW}s, {\small EMW}s and a
background plasma flow should occur also in less idealized situations than
considered in this paper, but in a more involved way. An improved model that
correctly describes the long term evolution of the interaction between a
{\small GW}, an {\small EMW} and the background flow in an inhomogeneous
background is a project for future research.


\begin{thebibliography}{99}
\bibitem{Gertsenstein}M. E. Gertsenshtein, Zh. Eksp. Teor. Fiz. \textbf{41},
113 (1961) [Sov. Phys. JETP, \textbf{14}, 84 (1962)].

\bibitem{Zeldovich}Y. B. Zeldovich, Zh. Eksp. Teor. Fiz. \textbf{65}, 1311
(1973) [Sov. Phys. JETP, \textbf{65}, 1311 (1973)].

\bibitem{Grishchuk}L. P. Grishchuk and A. G. Polnarev, in \emph{General
Relativity and Gravitation}, edited by A. Held (Plenum, New York, 1980) Vol. 2.

\bibitem{Radiowave}M. Marklund, G. Brodin and P. Dunsby, Astrophys. J.
\textbf{536}, 875 (2000).

\bibitem{Phacc}G. Brodin, M. Marklund and M. Servin, Phys. Rev. D 63, 124003
(2001);\newline M. Maggiore, Phys. Rep. \textbf{331}, 283 (2000);\newline Yu.
G. Ignat'ev, Phys. Lett. A \textbf{320}, 171 (1997).

\bibitem{Alfvénwave}M. Servin, G. Brodin, M. Marklund and M. Bradley, Phys.
Rev. E \textbf{62}, 8493 (2000).

\bibitem{Papadopulos}D. Papadopoulos, N. Stergioulas, L. Vlahos, et al. A \& A
\textbf{377}, 701 (2001).

\bibitem{SN}R. Bingham \emph{et al.}, Phys. Scr. \textbf{T75}, 61 (1998).

\bibitem{Kleidis}K, Kleidis, H. Varvoglis and D. Papadopoulos, Class. Quantum
Grav. \textbf{13}, 2547 (1996).

\bibitem{Weiland}J. Weiland and H. Wilhelmsson, \emph{Coherent Nonlinear
Interaction of Waves in Plasmas} (Pergamon Press, New York, 1977);\newline A.
Craik, \emph{Wave interactions and fluid flows} (Cambridge Univeristy Press 1985).

\bibitem{Anile}A. M. Anile, J. K. Hunter and B. Turong, J. Math. Phys.
\textbf{40}, 4474 (1999).

\bibitem{Greco}A. Greco and L. Seta, Class. Quantum Grav. \textbf{15}, 3655 (1998).

\bibitem{Macedo}P. G. Macedo and H. Nelson, Phys. Rev. D \textbf{28}, 2328 (1983).

\bibitem{Algebraic}In general, the source terms may contain derivatives on
$f_{(n)}$ and $f_{(m)}$. As we restrict ourself to weak interaction these
derivatives are later approximated by algebraic expressions, i.e. involving
frequencies and wave numbers.

\bibitem{source terms}M. Marklund, G. Brodin and P. K. S. Dunsby, Astrophys.
J. \textbf{536}, 875 (2000);\newline G. Brodin, M. Marklund, P. K. S. Dunsby,
Phys. Rev. D \textbf{62}, 104008 (2000).

\bibitem{Isacson}R. A. Isaacson, Phys. Rev. \textbf{166}, 1263 (1968).

\bibitem{Plasmawaves}T. H. Stix, \emph{Waves in Plasmas} (American Institite
of Physics, New York, 1992);\newline D. R. Nicholson, \emph{Introduction to
Plasma Theory} (John Wiley \& Sons Inc., 1983).

\bibitem{Landau}L. D. Landau and E. M. Lifshitz, \emph{The Classical Theory of
Fields} (Pergamon Press, Oxford, 1975).

\bibitem{Chen}F. Chen, \emph{Introduction to Plasma Physics and Controlled
Fusion} (Plenum Press, New York, 1984).

\bibitem{efflin}The dispersion relation, however, depends on the wave
amplitude through the gamma factors, but due to the circular polarization the
gamma factors are constant (for free waves).

\bibitem{Stenflo}L. Stenflo, Phys. Scr. \textbf{14}, 320 (1976).

\bibitem{Flow}Formally, the drift velocities, $v_{z(i)}$, cannot be taken
arbitrarily. They should fulfill $j_{z}=0$ for the given solution to be valid,
or else the drifts produce a background magneic field that is not accounted
for in the solution. On the other hand, one can always divide a fluid
component into several species, each with a different drift velocity and
thereby regain freedom in the relative drifts. Thus we discard this
restriction because it should not influence the final result qualitatively.

\bibitem{Strictly}From a mathematical point of view, our large {\small EMW}
solution (together with the background drift) contains more energy than the
state with only the \emph{pure} drift flow. However, from energy conservation
it is clear that the electromagnetic field amplitude cannot be physically
altered without simultaneously affecting the drift velocity. When confirming
that the wave energy is negative, one should compare with the state where this
change in drift flow has been taken into account.

\bibitem{Velocity}One could expect that the curves in FIG 1. should be even
functions of the velocity. The symmetry is, however, broken by the definite
propagation direction. The important point is that the relation between the
wave polarization and the particle gyration (determined by $\Omega_{ce}$) is
fixed by the condition $\omega=k$, which depends on the flow velocities, as
seen by Eq. (\ref{B-relation}).

\bibitem{Phase}In general also the phase $\phi$ evolves in time. Eqs.
(\ref{real1})-(\ref{real3}) are still valid but should be supplemented with an
evolution equation for the phase, see e.g. Ref. \cite{Weiland}. For our
purposes, however, it suffices to consider the special cases where the phase
is constant.

\bibitem{Conversion rate}By \emph{conversion rate} we mean the inverse time
scale for one wave to be converted into another type of wave.

\bibitem{Mode conv.}R. Littlejohn and W. Flynn, Phys. Rev. Lett. \textbf{70},
1799 (1993).

\bibitem{BVPnote} This example makes use of the substitution 
$\partial_{t}\rightarrow v_{g}\partial_{z}$ appropriate for a boundary value 
problem. Effects of an inhomogeneous background is neglected.

\bibitem{Morou} G. A. Morou, C.P.J. Barty and M.D. Perry, Phys. Today, 
\textbf{51}, No 1, 22 (1998).
\end{thebibliography}
\end{document}